\documentstyle[sprocl]{article}
\def\Journal#1#2#3#4{{#1} {\bf #2}, #3 (#4)}
\def\NPB{{\em Nucl. Phys.} B}
\def\PLB{{\em Phys. Lett.}B}
\def\PRL{\em Phys. Rev. Lett.}
\def\PRD{{\em Phys. Rev.} D}

\def\ra{\rightarrow}
\def\al{\alpha}
\def\be{\begin{equation}}
\def\ee{\end{equation}}

\begin{document}
\title{EXCITED HEAVY MESONS FROM QCD SUM RULES}
\author{ Chun Liu}
\address{Center For Theoretical Physics, Seoul National University, Seoul
151-742, Korea}

\maketitle
\abstracts{Orbitally excited $L=1$ charm mesons are studied 
by QCD sum rules in the framework of heavy quark effective 
theory.  The meson masses and the strong decay widths are 
obtained.  This talk is based on our works of refs. 1 and 2 collaborated
with Y.B. Dai, C.S. Huang, M.Q. Huang and H.Y. Jin.}

The study of the excited heavy mesons are interesting for the following 
reasons~\cite{falk}.  First, it may help us understand QCD more deeply in the 
nonperturbative aspect.  Second, it is an alternative application of the 
heavy quark effective theory (HQET), in addition to that of ground state 
heavy mesons.  Finally, it is particularly useful for tagging in CP experiments
~\cite{eichten}.\\

In this talk, we focus on the $L=1$ excited charm mesons.  
It is heuristic to illustrate them in the following way, although it is not
accurate ($L$ is not a conserved quantity).
For the ground 
state ($L=0$) mesons which we are familiar with, there are two states with the
$J^P$ quantum numbers: $D(0^-)$ and $D^*(1^-)$.  Their difference only lies 
in the heavy quark spin direction.  For the $L=1$ case, there are four 
states: $D_0^*(0^+)$, $D_1'(1^+)$, $D_1(1^+)$ and $D_2^*(2^+)$.  In the 
constituent picture, $D_0^*$ contains one heavy quark and one light 
quark with same spin direction which is opposite to that of the orbital 
angular momentum.  $D_1'$ is the same as $D_0^*$ up to a heavy quark spin 
flip.  $D_1$ is different from $D_0^*$ by the light quark spin flip.  And 
$D_2^*$ is the same as $D_1$ except for the heavy quark spin direction.\\

These kinds of heavy mesons can be systematically studied 
by HQET~\cite{hqef}.  In the heavy quark limit, 
the heavy quark SU(2) spin symmetry implies that the four excited heavy
mesons can be grouped into two doublets ($D_0^*$, $D_1'$) and 
($D_1$, $D_2^*$)~\cite{strong}.  It is interesting to note that although 
$D_1'$ and $D_1$ have same quantum numbers, they are clearly separated in the 
heavy quark limit.  Their mixing is at the order of $1/m_c$, which can be 
studied in HQET.  In the following, we present the HQET sum rule calculations 
for masses and strong decays of the excited charm mesons in the leading 
order.\\

The QCD sum rule~\cite{qcdsr} is a nonperturbative method rooted in QCD 
itself.  
The relevant interpolating 
current should be fixed.  For the excited heavy mesons, we wrote the 
currents as follows~\cite{dai1},  
\be
J=\frac{1}{\sqrt{2}}\bar{h}_v\Gamma D_{\rho} q~,
\ee
with $\Gamma$ denoting some $\gamma$ matrices.
$h_v$ is the heavy quark field with velocity $v$ in HQET.
We find that
\begin{equation}
\begin{array}{llll}
\Gamma&=&-\gamma_t^{\rho} ~~~&{\rm for}~~~ D_0^*~,\\
\Gamma&=&\gamma_5\gamma_t^{\mu}\gamma_t^{\rho} ~~~&{\rm for}~~~ D_1'~,\\
\Gamma&=&\displaystyle -\sqrt{\frac{3}{2}}\gamma_5(g_t^{\mu\rho}-\frac{1}{3}
\gamma_t^{\mu}\gamma_t^{\rho}) ~~~&{\rm for}~~~ D_1~,\\[3mm]
\Gamma &=&\displaystyle \frac{1}{2}(\gamma_t^{\mu}g_t^{\nu\rho}+
\gamma_t^{\nu}g_t^{\mu\rho})-\frac{1}{3}g_t^{\mu\nu}\gamma_t^{\rho}~~~ 
&{\rm for}~~~ D_2^*~,\\[3mm]
\end{array}
\end{equation}
where $\gamma_t^{\mu}\equiv\gamma^{\mu}-v^{\mu}\!\not\! v$ (In the rest frame 
of the meson, $\gamma_t^{\mu}=(0, \vec{\gamma}$)) and 
$g_t^{\mu\nu}\equiv g^{\mu\nu}-v^{\mu}v^{\nu}$.  We have shown that 
the current for one of the two $1^+$ state does not couple to the other
in the limit $m_Q\ra\infty$.  The related decay constant 
$f$ is defined as 
\be
\langle 0|J^{\dag}|D^{**}(v, \eta)\rangle \equiv f\eta~,
\ee
where $D^{**}$ stands for any of the four excited mesons with polarization 
vector $\eta$, which is the ground state of $J$ (For more details, see ref. 
1).\\

\section{ Masses }
In HQET, the meson masses are expanded as
\be
M=m_Q
+\bar{\Lambda}
+O(\frac{1}{m_Q})~,\\[3mm]
\ee
where $\bar{\Lambda}$ denotes the meson masses defined in HQET.  They are 
independent of heavy quark flavor.  For the sake of calculating them, the 
Green's function is written as
\be
\Gamma(\omega)=i\int d^4x e^{ik\cdot x}\langle 0|{\rm T}J^{\dag}(x)J(0)
|0\rangle~,~~~\omega=2k\cdot v~.
\ee
It can be expressed in the hadronic language,
\be
\Gamma(\omega)=\frac{2f^2}{2\bar{\Lambda}-\omega}+{\rm resonance}~.
\ee
The 
Feynman diagrams can be found in ref. 1.  The condensates are included up to 
dimension five.
Here the usual duality hypothesis is used for the resonance contribution.
After Borel transformation, the sum rules for $\bar{\Lambda}$'s are obtained 
then
\begin{equation}
\begin{array}{lll}
\displaystyle\bar{\Lambda}_{\left(\begin{array}{c}0^+\\1^+\end{array}\right)}
&=&\displaystyle\frac{\frac{3}{8\pi^2}\int^{\omega_c}_0d\nu\nu^5 e^{-\nu/T}}
{\frac{3}{4\pi^2}\int^{\omega_c}_0d\nu\nu^4 e^{-\nu/T}
-m_0^2\langle\bar{q}q\rangle}~,\\[3mm]
\displaystyle\bar{\Lambda}_{\left(\begin{array}{c}1^+\\2^+\end{array}\right)}
&=&\displaystyle\frac{\frac{1}{32\pi^2}\int^{\omega_c}_0d\nu\nu^5 e^{-\nu/T}
-\frac{1}{16\pi^2}\langle\al_sGG\rangle T^2}
{\frac{1}{16\pi^2}\int^{\omega_c}_0d\nu\nu^4 e^{-\nu/T}
-\frac{m_0^2}{3}\langle\bar{q}q\rangle
-\frac{T}{8\pi} \langle\al_sGG\rangle }~.\\[3mm]
\end{array}
\end{equation}
Imposing usual criterium for the upper and lower bounds of the Borel 
parameter $T$, we obtain the numerical results,
\be
\bar{\Lambda}_{\left(\begin{array}{c}0^+\\1^+\end{array}\right)}=0.90\pm 0.10
~{\rm GeV}~, ~~~
\bar{\Lambda}_{\left(\begin{array}{c}1^+\\2^+\end{array}\right)}=0.95\pm 0.10
~{\rm GeV}~,
\ee
where the error comes from the uncertainty of $\omega_c\sim 0.5$ GeV.\\

\section{ Strong Decays }
The pionic decays of heavy hadrons can be studied by combing heavy quark 
symmetry and chiral symmetry~\cite{hhcl}.  The excited heavy meson 
decays have been studied in this framework~\cite{luke}$^,$~\cite{koerner}.  
We will not use soft pion approximation however, because the pion energy is 
about 500 MeV in this case.  The decay widths are given by, {\it e.g.},
\begin{equation}
\begin{array}{lll}
\Gamma(D_0^*\ra D\pi)&=&\displaystyle\frac{3}{8\pi}g'^2|\vec{p_{\pi}}|~,
\\[3mm]
\Gamma(D_2^*\ra D\pi)&=&\displaystyle\frac{1}{20\pi}g^2|\vec{p_{\pi}}|^5~, 
\\[3mm]
\end{array}
\end{equation}
where sum over charged and neutral pion final states has been implied, and 
$g'$ and $g$ are universal quantities describing the decay amplitudes
which we are going to calculate by QCD sum rules.\\

For this purpose, the Green's function is constructed as
\be
\tilde{\Gamma}=i\int d^4x e^{-ik\cdot x}\langle \pi(q)|{\rm T}J_{D^{(*)}}(x)
J^{\dag}(0)|0\rangle~,
\ee
with $J_{D^{(*)}}$ being the currents for $D^{(*)}$ mesons.  
The hadronic expression is 
\be
\tilde{\Gamma}=\frac{g^{(')} f_{D^{(*)}}f_{D^{**}}}
{(2\bar{\Lambda}_{D^{(*)}}-2v\cdot k)(2\bar{\Lambda}_{D^{**}}-2v\cdot k')}
+\frac{c}{2\bar{\Lambda}_{D^{(*)}}-2v\cdot k}
+\frac{c'}{2\bar{\Lambda}_{D^{**}}-2v\cdot k'}+{\rm res.}~,\\[3mm]
\ee
where $k'=k-q$.  Instead of taking soft pion limit, we put 
$2v\cdot (k-k')=2(\bar{\Lambda}_{D^{**}}-\bar{\Lambda}_{D^{(*)}})$.
In the operator product expansion of HQET, 
\be
\tilde{\Gamma}=i\int_0^{\infty}dt e^{\frac{\omega t}{2}}\frac{1}{\sqrt{2}}
\langle\pi(q)|{\rm T}\bar{q}(vt)\Gamma_{D^{(*)}}\frac{1+\not\! v}{2}\Gamma
D_{\rho}q(0)|0\rangle ~.
\ee
What we need is $\langle\pi|\bar{q}^a_{\al}(x)D_{\rho}q^b_{\beta}|0\rangle$
which can be found in ref. 2.  With
\begin{equation}
\begin{array}{l}
\displaystyle 
g_2=\frac{f_{\pi}}{4}~, h_1=0~, 
h_2=-\frac{1}{12f_\pi}\langle\bar{q}q\rangle~,
a_1=-\frac{m_0^2}{8f_{\pi}}\langle\bar{q}q\rangle~, 
c_1=\frac{1}{36}f_{\pi}m_1^2~,\\[3mm]
\displaystyle
c_2=-\frac{1}{24f_\pi}\langle\bar{q}q\rangle~,
b_1=\frac{a_1}{3}~, d_1=0~, 
d_2=-\frac{e_2}{3}+\frac{1}{12f_\pi}\langle\bar{q}q\rangle~,
m_1^2\simeq 0.2 {\rm GeV}^2~,\\[3mm]
\end{array}
\end{equation}
and $e_2$ which is defined as 
\be
e_2(n\cdot q)^3=\displaystyle \frac{i}{2}\langle\pi^i(q)|\bar{q}
\frac{\tau_i}{2}\gamma_5\not\! n(n\cdot D)^2q|0\rangle~,~~~n^2=0~,\\[3mm]
\ee
and is fixed from QCD sum rule~\cite{dai2}:
$e_2\simeq -0.015\pm 0.002~{\rm GeV}$,
the final sum rules for $g'$ and $g$ are
\begin{equation}
\begin{array}{lll}
\displaystyle 
g'f_{D_0^*}f_D&=&\displaystyle
2\{\bar{\Lambda}_{D_0^*}(3h_1+3h_2\Delta-g_2\Delta^2)-[-3b_1
+3\Delta(c_1 +d_1)\\[3mm]
&&\displaystyle +\Delta^2(4c_2+2d_2-e_2)](1+\frac{2\bar{\Lambda}_{D_0^*}}{T})\}
e^{\frac{2\bar{\Lambda}_{D_0^*}}{T}}~,\\[3mm]
\displaystyle 
gf_{D_2}f_D&=&\displaystyle
2\left[\bar{\Lambda}_{D_2}g_2+(c_2-e_2\Delta-d_2)(1
+\frac{2\bar{\Lambda}_{D_2}}{T})\right]e^{\frac{2\bar{\Lambda}_{D_2}}{T}}~,
\\[3mm]
\end{array}
\end{equation}
where $\Delta=\bar{\Lambda}_{D^{**}}-\bar{\Lambda}_D$.  The range of the 
Borel parameter is fixed by requiring higher $\frac{1}{T}$ term to be small 
in OPE for original sum rules, and $T$ being not too large than 
$2(\bar{\Lambda}^1-\bar{\Lambda})$ where $\bar{\Lambda}^1$ is the 
mass of first radial excitation in HQET.  We obtain the numerical results
\begin{equation}
\begin{array}{llll}
g'&\simeq& 1.7\pm 0.5~,\\
g &\simeq& 4.6\pm 1.0 ~~~{\rm GeV}^{-2}~.\\
\end{array}
\end{equation}

For discussion, our methods improve the calculations by sum rules in full 
QCD~\cite{col} which have problems of including contamination of the other
$1^+$ state, and using soft pion approximation.  
The $1/m$ corrections~\cite{dai3} and radiative
corrections can be included systematically.  The experiment data on $D_2^*$
width~\cite{pdg} can be explained by assuming the pionic decay mode dominant. 
However, that of $D_1$ cannot be understood in this way.  It maybe due to
the mixing of two $1^+$ states at the order of $1/m$ as well as that  
$D_1\ra D^{(*)}\rho$ have relatively large branching ratios.

\section*{Acknowledgments}
I am grateful to Professors Yuanben Dai and Chaoshang Huang and Drs Mingqiu
Huang and Hongying Jin for instructive and enjoyable collaboration.  This 
work was supported in part by KOSEF (SRC).

\end{document}